# Electromagnetic properties of polycrystalline diamond from 35K to room temperature and microwave to terahertz frequencies

**Jean-Michel Le Floch[1,2,*], Romain Bara[1,2], John G. Hartnett[1], Michael E. Tobar[1,2]**

[1] School of Physics, University of Western Australia, 35 Stirling Hwy, 6009 Crawley, Australia

[2] Australian Research Council - Centre of Excellence Engineered Quantum Systems (EQUS)

Emails: lefloch@cyllene.uwa.edu.au, rbara@cyllene.uwa.edu.au, john@physics.uwa.edu.au, mike@physics.uwa.edu.au

**David Mouneyrac[1,3], Damien Passerieux[3], Dominique Cros[3]**

[3] XLIM, UMR CNRS Université de Limoges n°6172, 123 avenue A. Thomas, 87060 Limoges Cedex, France

Emails: davmoun@cyllene.uwa.edu.au, damien.passerieux@xlim.fr, Dominique.cros@xlim.fr

**Jerzy Krupka[4]**

[4] Institute of Microelectronics and Optoelectronics, Department of Electronics, Warsaw University of Technology, Koszykowa 75, 00-662 Warsaw, Poland

Email: j.krupka@imio.pw.edu.pl

**Philippe Goy[5,6], Sylvain Caroopen[5]**

[5] Laboratoire Kastler-Brossel, Département de Physique de l'Ecole Normale Supérieure, 24 rue Lhomond, 75005, Paris, France

---

[*] Corresponding author

[6] AB MILLIMETRE, 52 rue Lhomond, 75005 Paris, France

Emails: philippe.goy@abmillimetre.com, sylvain.caroopen@abmillimetre.com

**Abstract:**

Dielectric resonators are key components for many microwave and millimetre wave applications, including high-Q filters and frequency-determining elements for precision frequency synthesis. These often depend on the quality of the dielectric material. The commonly used material for building the best cryogenic microwave oscillators is sapphire. However sapphire is becoming a limiting factor for higher frequencies design. It is then important to find new candidates that can fulfil the requirements for millimetre wave low noise oscillators at room and cryogenic temperatures. These clocks are used as a reference in many fields, like modern telecommunication systems, radio astronomy (VLBI), and precision measurements at the quantum limit. High-resolution measurements were made of the temperature-dependence of the electromagnetic properties of a polycrystalline diamond disk at temperatures between 35 K and 330 K at microwave to sub-millimetre wave frequencies. The cryogenic measurements were made using a $TE_{01\delta}$ dielectric mode resonator placed inside a vacuum chamber connected to a single-stage pulse-tube cryocooler. The high frequency characterization was performed at room temperature using a combination of a quasi-optical two-lens transmission setup, a Fabry-Perot cavity and a whispering gallery mode resonator excited with waveguides. Our CVD diamond sample exhibits a decreasing loss tangent with increasing frequencies. We compare the results with well known crystals. This comparison makes clear that polycrystalline diamond could be an important material to generate stable frequencies at millimetre waves.

# 1. Introduction

Good quality dielectric resonators are very important for modern telecommunication systems and are required for cellular phone base stations and low phase noise and highly frequency stable oscillators. In general the performance of microwave and millimetre wave devices often depends on the quality of the dielectric materials used. The Q-factor of a standard dielectric resonator is usually limited by the dielectric loss tangent of the material. To date the best microwave oscillators have been built based on sapphire resonators.[1-3] However sapphire is not so low loss in the extremely high frequency (EHF) band due to the loss tangent $f^1$ dependence caused by the tail of the first phonon resonance between 10 and 20 THz.[4] Other materials such as Teflon and quartz do not have such a frequency dependence. Nevertheless these latter are not suitable either for the EHF band applications. Because of its unique combination of thermal and chemical stability, low thermal expansion and high optical transparency over a wide spectral range, synthetic diamond is becoming a popular material for optical applications.[5] Diamond wafers are also of some interests for its heat dissipating capabilities such high power switches at power stations, high-frequency field-effect transistors and light-emitting diodes.[6-9] Diamond has not previously been of interest for making dielectric resonators in the transition band from microwave to terahertz due to various technical problems including heavy cost, secrecy of manufacturing, long and tough growth process for reasonable size sample for microwave to millimetre wave domain.

In this work we investigate the properties of a single polycrystalline synthetic diamond sample. Polycrystalline diamond is easier to source and a cheaper alternative compared to the crystalline material. Here we present results of our characterization of a sample, 1.227 ± 0.001 mm thick with a 10.086 ± 0.001 mm diameter, over a very broad frequency and temperature range. The sample was obtained from Diamond Materials GmbH made by chemical vapour deposition (CVD). We show that the loss tangent decreases with increasing frequencies and is of the order of $10^{-4}$ above 500 GHz. Also the temperature coefficient of permittivity is given from liquid helium to room temperatures. These results show that the material is suitable for constructing high-Q dielectric resonators based on trapped modes[10-11] at these frequencies and could find application in the generation of low phase noise and ultra stable signals.

## 2. Measurements techniques

In order to cover a wide range of frequencies and temperatures, it has been necessary to split the measurements in different stages using different setups suitable for the frequency range we are investigating. We use four different measurement techniques, including the whispering gallery modes and the transverse electric modes techniques, the quasi-optical free space bench from AB Millimetre and the Fabry-Perot cavity. Some of the measurements can only give a close approximation of the dielectric properties due to their broad band settings and some others are working at a particular resonance frequency which gives the highest precision in the intrinsic crystal properties determination. However, in practice it is difficult to use some of these previous single resonance frequency measurement techniques on small size samples like the diamond we investigated, due to problems with excitation of the resonance at very high frequencies.

One of the most accurate techniques for the characterization of the dielectric properties of low-loss crystalline and polycrystalline materials is the whispering gallery mode (WGM) technique, which has been used very successfully from the microwave to the millimetre waves.[12-14] The technique uses higher order hybrid modes excited in a cylindrical or spherical dielectric. The WGM technique is especially useful for measurements of extremely low-loss materials at both cryogenic[15,16] and room temperature. The relationship between resonant frequencies dimensions of the resonant cavity, and permittivity of the sample under test is calculated with a rigorous electromagnetic simulation software based on the method of lines (MoL)[17] with an accuracy better than $10^{-4}$. The total uncertainty in permittivity using WGM was estimated to be less than 0.05%. The uncertainty was mainly limited by uncertainty in sample dimensions measurements. Experimental and calculated resonant frequencies of WGM differed by no more than 0.01%. In practice it is important to identify only the first few modes belonging to the WGM family because experimental identification of proper modes becomes extremely difficult with increasing frequency due to the increase in the density of spurious modes.

The WGM family is varying linearly with the increasing azimuthal mode number which helps to predict and select the right frequency peak.

The TE$_{01\delta}$ mode has also been traditionally used for precise measurements and characterization of dielectric materials.[18] Here we use this technique to characterize the dielectric loss tangent and permittivity as a function of frequency in the lower microwave frequency range. The resonant structure used for measurements, along with the electric field density plot of the mode is shown in figure 1. Accurate complex permittivity determination using the quasi TE$_{01\delta}$ mode requires numerical electromagnetic computations. For the structure shown in figure 1 this can be done by employing one of several rigorous techniques, e.g. finite element,[19] finite difference (method of lines)[17] or mode matching.[20] The uncertainties based on this technique lay also on the uncertainty of the sample dimensions measurements. One reason why real permittivity is rarely measured using this technique is the difficulties in accurate electromagnetic modelling of the TE$_{01\delta}$ mode resonator structure. With increasing frequency the dimensions of the samples become very small and measurement uncertainties increase. This is the main reason why this technique is rarely used for measurements at frequencies above 30 GHz. Comparing typical deviations between WG and TE$_{01\delta}$ modes characterisation methods in retrieving the permittivity values are within 0.1%.[18] The loss tangent values for both above mentioned techniques are limited by the uncertainty of the resonance bandwidth within 10%.

To measure higher frequency characteristics, WGM cavities were coupled using standards metallic waveguides (W/G) over the 50 to 61 and 111 to 281 GHz frequency bands in a transmission configuration. At those frequencies coupling with coaxial probes is very complex to achieve due to their very small sizes. The use of W/G simplifies the experiment setup. The series of measurements has been done in a closed copper cylindrical cavity. The loss tangent of the dielectric sample in this configuration may be limited with the surface resistance of the cavity walls. At even higher frequencies of 672-1000 GHz two techniques were used with consistent results. 1) The quasi-optical two-lens transmission setup, as shown in figure 2a, employed the MVNA-8-350 vector analyser from AB MILLIMETRE. The sample was placed at the point of minimum beam waist and behaves as an infinite plane parallel Fabry-Perot resonator, which allowed very wide frequency sweeps. 2) The perturbation cavity technique, as shown in figure 2b (described below), where the dielectric sample makes a perturbation of a high-Q semi-spherical Fabry-Perot metallic resonator tuned at chosen fixed frequencies. The loss tangent limit of both above mentioned millimetre wave techniques is of the order of $10^{-4}$ and an accuracy of $10^{-3}$ of the permittivity value.[21]

To measure the temperature dependence, the fundamental TE mode family was chosen. Modes were simulated using method of lines and mode matching software to predict the frequency ($Fr$), the geometric factor ($G$), and the electric energy filling factor ($pe$). Mode Q-factor and frequency measurements were made as a function of temperature from the transmission coefficient of the diamond-loaded cavity, as shown in figure 1. Modes were weakly coupled via coaxial cables and loop probes so the couplings were much less than unity. The cavity was attached to the cold finger of a single stage pulse tube cryocooler.[2] Simulations were compared with these measurements for the fundamental mode at 22.3 GHz. The loss tangent was calculated using,

$$\tan\delta = \frac{1}{pe}\left(\frac{1}{Q_0} - \frac{R_S}{G}\right) \quad , \tag{1}$$

where $G$ and $pe$ are defined as following:

$$G = \frac{\omega \iiint_V \mu_o H H^* dV}{\oiint_S H_t H_t^* dS} \quad , \tag{2}$$

$$pe = \frac{We_{material}}{We_{Total}} = \frac{\iiint_{V_{material}} \varepsilon_{material} E E^* dV}{\iiint_V \varepsilon(v) E E^* dV} \quad . \tag{3}$$

Here $Q_0$ is the unloaded Q-factor and $R_S$ is the surface resistance of the enclosing metal cavity. The coupling to the cavity was such that the unloaded Q-factor was approximately equal to the loaded Q-factor. In these formulas, $V$ is the total volume of the resonator, $V_{material}$ is the volume of the material and $S$ is the surface of the cavity walls.

The determination of the unloaded Q-factor ($Q_o$) from the Whispering Gallery and TE modes technique can be deduced from $\beta_{1,2}$ the coupling coefficients on both ports and the loaded Q-factor ($Q_L$) we measure. They are given by the relationship between reflection ($S_{11}$) and transmission ($S_{21}$) measurements as following:

$$\beta_1 = -\frac{S_{21}^2}{S_{21}^2 + S_{11}^2 - 1} \text{ and } \beta_2 = -\frac{S_{11}^2 - 2S_{11} + 1}{S_{21}^2 + S_{11}^2 - 1} \tag{4}$$

$$Q_o = (1 + \beta_1 + \beta_2)Q_L \tag{5}$$

However it is not the only technique to determine the unloaded Q-factor as it has been suggested and extended in Fittipaldi et al paper.[22]

## 3. Results

The coupling to the cavity was such that the unloaded Q-factor was approximately equal to the loaded Q-factor. The cavity transmission loss was set to at least 36 dB in the $TE_{01\delta}$ cavity. Note that electric energy filling factor in the Teflon support post is very small ($2 \times 10^{-4}$) and can be ignored in the calculations. The resulting real part of the permittivity and the loss tangent as a function of temperature are shown in figures 3 and 4.

The temperature dependence of the sample permittivity at 22.3 GHz was determined from measurements (Table I) to have the following form:

$$\varepsilon(T) = 5.706 - 1.687 \times 10^{-4} T + 2.365 \times 10^{-7} T^2 \tag{6}$$

It has been proven[11] that computational errors in most practical cases are smaller than measurement errors and it has already been shown for a simple resonant structure, the computational error increases for higher order modes. One can see that computational errors are about 0.1% for most of the quasi $TE_{0n\delta}$ modes.

Note that temperature dependence of the loss tangent, as shown in figure 4, can be described by the following piecewise equation:

$$\tan \delta(T) = 2.137 \times 10^{-4} T^{0.329} \quad for \quad T > 100K \tag{7}$$

$$\tan \delta(T) = 7.950 \times 10^{-4} T^{0.045} \quad for \quad T < 100K \tag{8}$$

For dielectric loss tangents that are of the order $10^{-4}$, the measurement uncertainties on the resonance bandwidth are about 10%.

The permittivity of the sample was also determined using whispering gallery modes from the X to V band to investigate possible frequency dependence. WG modes were simulated using the method of lines and the permittivity calculated from their measured resonance frequencies. The modes were excited in the cavity using rectangular waveguides to couple in the high frequency interrogation signal. The results are shown in figure 5. The loss tangent was estimated from the half power bandwidth of the weakly coupled resonance and was of the order $10^{-3}$ at room temperature. Due to the energy confinement in the resonator, the metallic losses are neglected.

We also investigated the permittivity ($\varepsilon_r$) or refractive index ($n$) ($n = \sqrt{\varepsilon_r}$) over a higher frequency range (above 600 GHz) using the open-air cavity set up in figure 2(a). The relationship between the periodicity of the transmission maxima ($\Delta Fr$) and the refractive index (n) is as following:

$$n = \frac{c}{2 \times \Delta Fr \times s} \tag{9}$$

where $s$ is the thickness of the sample. This expression is broadband, i.e. gives the refractive index over the range of frequencies studied. The relationship between the phase rotation (t) downwards from the cavity "empty" to "loaded" and the refractive index (n) is defined as:

$$(n-1)s = t\lambda \tag{10}$$

But this expression is valid at <u>only one</u> frequency. The results from using the open air cavity characterization method are determined from a point to point calculation and an automatic least mean squares fitting software built into the AB MILLIMETRE MVNA-8350 network analyser at frequencies from 672 GHz and 1,000.6 GHz. First we have used the broadband approach (Eq. 9) which gives an averaged value of the refraction index over the frequency interval between two peaks to obtain the order of the permittivity value. Then we compared it with a single frequency point characterisation using the phase rotation (Eq. 10). The latter is the method we retained for our data as it is more precise. From this we determined the relative permittivity, given by,

$$\varepsilon^* = \varepsilon'(1 - j\tan\delta) = 5.661\,(1 - j6.3\times10^{-5}) \tag{11}$$

In order to verify this result we used a Fabry-Perot cavity, from which we extracted the data as follows. The relationship between the phase rotation ($t$) and the refractive index ($n$) here is given by,

$$\frac{(n-1)s}{\lambda} = \frac{\Delta\varphi}{2\pi} + k \tag{12}$$

where $\lambda$ is the wavelength at resonance and $k$ is an integer number of turns. $\Delta\varphi$ gives the phase rotation, which corresponds to less than one turn. The loss tangent (tan$\delta$ is obtained from the damping of the transmitted signal and is given by[21]

$$\tan\delta = \frac{1.1\,\alpha}{nFr} \tag{13}$$

where α, the attenuation, is in units of dB/cm and the resonance frequency $Fr$ is in GHz units. The parameter α is determined from measuring the loss at the transmission maxima. The loss tangent at these frequencies was determined to be of the order $10^{-5}$.

These results are all combined from the different methods and shown in figure 5, to see the general trend of the real part of the permittivity on frequency. The broken line is a power law fit, which shows very little dependence on frequency over the range 22 GHz to 1 THz, and is given by $\varepsilon(f) = 5.67 \times f^b$ where $f$ is in GHz units and b = $(9\pm10) \times 10^{-4}$. Detailed analysis of uncertainties in permittivity evaluations lead to the conclusion that the most probable error due to the scatter in the data is due to numerical computations and dimensional measurement errors of the apparatus and sample.

## 4. Comparison with well known crystals and other investigations

Our main interest in this diamond characterisation is to find a possible candidate for the next generation of ultra stable low phase noise oscillators. That means the crystal has to be as low loss as possible to ensure a high spectral purity. Also it is crucial for such devices to find a material with a zero or very small sensitivity to temperature fluctuations. In figures 6 and 7, we compare the diamond sample with other crystals like sapphire[23] currently used in the best record ultra stable clocks and quartz, YAG[24-25], alumina[11]. In figure 6, it is really important to notice the loss tangent of diamond has an opposite sign with standard crystals. There is a significant improvement of the loss tangent at low temperatures and also at extremely high frequencies (EHF). The consistency of the data with previous studies at lower or same frequency bands[26-29] with smaller samples has been unexpected. The growth of a big bulk sample without creating a structure disorder has been successful. Otherwise it would have prevented the inversion of the loss mechanism.[30] This is the crucial element for a millimetre wave high Q dielectric resonator design. None of the low loss crystals we knew so far in a metrological domain is able to compete with such a crystal. For technical reasons we have not been able to proceed to temperature measurements at EHF but it is expected to benefit both conditions to achieve lower the loss tangent. In figure 7, it is shown the relative change of permittivity normalised to room temperature permittivity value. It is obtained by dividing the permittivity value for each temperature point with the room temperature permittivity value. The diamond temperature coefficient is as low as -10ppm/K. This is about seven times less temperature sensitive than sapphire for a same large temperature window from 150K to room temperature. However the temperature sensitivity at liquid Helium temperature (4.3K) is not as good as sapphire but it stays small enough to make oscillators.[31] As it is shown in figures 6 and 7, the CVD diamond sample we have characterised presents a much better loss mechanism than any other crystal known today and exhibits a low temperature drift coefficient. This means we are able to make a very high Q dielectric resonator in millimetre wave and build an oscillator with low sensitivity to temperature at room temperature which could be reduce either with a compensation mechanism or a control loop at cryogenic temperature. Thus it places this poly crystalline sample to a very best position as potential candidate to build the next millimetre wave oscillators.

## 5. Conclusion

In conclusion, the real part of the diamond permittivity does not show any frequency dependency as it is within the standard error of the fit. However the frequency dependence of loss tangent is very interesting as it improves with frequency which is atypical for a dielectric crystal. Also the improvement of its losses in decreasing temperature and its low temperature coefficient makes the diamond a very interesting candidate for building ultra stable millimetre wave oscillators, cryogenic high power amplifiers. It then makes possible to fit the VLBI requirements at EHF. It also allows new test of physics using the millimetre wave frequency band.

**Acknowledgements**:

This work was supported by the Australian Research Council (ARC) under a Discovery Project scheme and the V-band equipment has been purchased with a UWA Research Development Award. We are grateful to Tess Williams, Hua Li and Nicholas Bannan for their useful discussions. This research was jointly supported under Australian Research Council's funding schemes: Discovery Project project number DP0878108, Laureate Fellowship project number FL0992016 and Centre of Excellence Engineered Quantum Systems project number CE11E0082.

Table I: The resonance frequency (TE$_{01\delta}$), the electric energy filling factor ($pe$) in the diamond sample and in the Teflon supports and the $Q$-factor including losses due to the metal conductor of the resonant structure and the calculated geometric factor ($G$) as a function of temperature.

| T [K] | Fr meas. | Q | Loss [dB] | $\varepsilon$ | tan$\delta$ ($10^{-4}$) | Filling Factor ($pe$) | | G |
|---|---|---|---|---|---|---|---|---|
| | | | | | | Diamond | Teflon | |

|  | [GHz] |  |  |  |  |  | (10⁻⁴) |  |
|---|---|---|---|---|---|---|---|---|
| 35 | 22.279 | 1244 | 39.3 | 5.701 | 9.35 | 0.8323 | 1.94 | 590 |
| 40 | 22.279 | 1232 | 39.3 | 5.700 | 9.43 | 0.8325 | 1.94 | 590 |
| 50 | 22.279 | 1222 | 39.5 | 5.699 | 9.49 | 0.8323 | 1.95 | 590 |
| 60 | 22.279 | 1205 | 39.8 | 5.697 | 9.61 | 0.832 | 1.95 | 590 |
| 80 | 22.279 | 1187 | 40.4 | 5.694 | 9.72 | 0.8316 | 1.96 | 590 |
| 100 | 22.277 | 1141 | 41 | 5.692 | 10.08 | 0.8314 | 1.97 | 590 |
| 120 | 22.276 | 1104 | 41.6 | 5.689 | 10.38 | 0.8311 | 1.97 | 590 |
| 140 | 22.274 | 1052 | 42.4 | 5.687 | 10.87 | 0.8308 | 1.98 | 590 |
| 160 | 22.271 | 1004 | 43.2 | 5.686 | 11.36 | 0.8306 | 1.99 | 589 |
| 180 | 22.269 | 966 | 44.1 | 5.684 | 11.78 | 0.8302 | 1.99 | 589 |
| 200 | 22.265 | 922 | 44.9 | 5.682 | 12.32 | 0.8299 | 2.00 | 589 |
| 220 | 22.262 | 902 | 45.6 | 5.681 | 12.55 | 0.8298 | 2.01 | 589 |
| 250 | 22.256 | 860 | 46.5 | 5.679 | 13.12 | 0.8291 | 2.02 | 589 |
| 294 | 22.243 | 825 | 47.4 | 5.679 | 13.55 | 0.8286 | 2.03 | 589 |
| 295 | 22.246 | 822 | 38.7 | 5.678 | 13.62 | 0.8287 | 2.04 | 588 |
| 310 | 22.244 | 859 | 37.5 | 5.676 | 12.93 | 0.829 | 2.05 | 588 |
| 330 | 22.238 | 747 | 37 | 5.677 | 14.96 | 0.8287 | 2.05 | 588 |

Figure 1: Density plot (modulus squared) of the tangential electric field component $E_t$ for the fundamental mode $TE_{01}$ calculated using the method of lines in a diamond loaded copper cavity. The 2mm diameter Teflon support the diamond sample (diameter $D_d$ = 10.086mm and height $H_d$ = 1.227mm) inside a copper cavity (diameter $D_c$ = 18.05mm and height $H_c$ = 5mm). The $TE_{01\delta}$ mode excitation achieved with side loop probes.

Figure 2a: Quasi-optical two-lens transmission setup – AB MILLIMETRE

Figure2b: Fabry-Perot cavity

Figure 3: Relative permittivity versus temperature

Figure 4: Loss tangent versus temperature

Figure 5: Permittivity versus frequency, determination achieved using different techniques depending on the frequency range 1) TE modes, 2) WG modes and 3) the combination of the quasi optical Fabry-Perot cavity techniques.

Figure 6: Frequency dependence of the loss tangent of diamond compared with extrapolated data from sapphire[22], quartz, YAG[23-24], alumina[11] and respectively from 1 to 1,000 GHz. The crosses show the data points we measured from the diamond sample. TE, WG, OC/FP mean the technique employed to determine the diamond properties. TE stands for Transverse Electric mode, WG stands for Whispering Gallery mode, QO/FP stand for Quasi – Optical setup and Fabry Perot respectively.

Figure 7: Temperature coefficient of permittivity (TcP) of our diamond sample compared with sapphire and quartz crystals